\documentclass[aps,preprint,preprintnumbers,superscriptaddress]{revtex4-2}
\usepackage{amsmath}
\usepackage{amssymb}
\usepackage{mathrsfs}
\usepackage{graphicx}
\usepackage{bm}
\usepackage{epstopdf}
\usepackage{bookmark}
\usepackage{color}
\begin{document}
\title{Formation mechanisms and fluorescence properties of carbon
dots in coal burning dust from coal fired power plants}

\author{Zhexian Zhao}\thanks{Z. Zhao and W. Zhang contributed equally to this work.}
\affiliation{School of Physics and Astronomy and Key Laboratory of Yunnan Provincial
Higher Education Institutions for Optoelectronics Device Engineering,
Yunnan University, Kunming 650091, China}

\author{Weizuo Zhang}\thanks{Z. Zhao and W. Zhang contributed equally to this work.}
\affiliation{School of Physics and Astronomy and Key Laboratory of Yunnan Provincial
Higher Education Institutions for Optoelectronics Device Engineering,
Yunnan University, Kunming 650091, China}

\author{Jin Zhang}\email{zhangjin@ynu.edu.cn}
\affiliation{School of Physics and Astronomy and Key Laboratory of Yunnan Provincial
Higher Education Institutions for Optoelectronics Device Engineering,
Yunnan University, Kunming 650091, China}
\affiliation{Yunnan Carbon Based Technology Co. Ltd., Kunming 650028 China}

\author{Yuzhao Li}
\affiliation{School of Physics and Astronomy and Key Laboratory of Yunnan Provincial
Higher Education Institutions for Optoelectronics Device Engineering,
Yunnan University, Kunming 650091, China}

\author{Han Bai}
\affiliation{School of Physics and Astronomy and Key Laboratory of Yunnan Provincial
Higher Education Institutions for Optoelectronics Device Engineering,
Yunnan University, Kunming 650091, China}
\affiliation{Department of Radiation Oncology,
Yunnan Tumor Hospital, Kunming 650106, China}

\author{Fangming Zhao}
\affiliation{Yunnan Huadian Kunming Power Generation Co. Ltd., Kunming 650308, China}

\author{Zhongcai Jin}
\affiliation{Zhejiang Huachuan Industry Group Co. Ltd., Yiwu 322003, China}

\author{Ju Tang}
\affiliation{Department of Physics, School of Electrical and Information
Technology, Yunnan Minzu University, Kunming 650504, China}

\author{Yiming Xiao}\email{yiming.xiao@ynu.edu.cn}
\affiliation{School of Physics and Astronomy and Key Laboratory of Yunnan Provincial
Higher Education Institutions for Optoelectronics Device Engineering,
Yunnan University, Kunming 650091, China}

\author{Wen Xu}\email{wenxu\_issp@aliyun.com}
\affiliation{School of Physics and Astronomy and Key Laboratory of Yunnan Provincial
Higher Education Institutions for Optoelectronics Device Engineering,
Yunnan University, Kunming 650091, China}

\author{Yanfei L\"{u}}\email{optik@sina.com}
\affiliation{School of Physics and Astronomy and Key Laboratory of Yunnan Provincial
Higher Education Institutions for Optoelectronics Device Engineering,
Yunnan University, Kunming 650091, China}

\date{\today}

\begin{abstract}
Carbon dots (CDs) shows great application potential with their unique
and excellent performances. Coal and its derivatives are rich in aromatic
ring structure, which is suitable for preparing CDs in microstructure.
Coal burning dust from coal-fired power plants can be utilized as
a rich resource to separate and extract CDs. It has been
shown in our results that there have two main possible mechanisms for
the formation of CDs in coal burning dust. One is the self-assembly
of polycyclic aromatic hydrocarbons contained in coal or produced by
incomplete combustion of coal. The other mechanism is that the bridge
bonds linking different aromatic structures in coal are breaking
which would form CDs with different functional groups when the coals
are burning at high temperature. Under violet light
excitation at 310-340 nm or red light at 610-640 nm, CDs extracted
from coal burning dust can emit purple fluorescence around 410 nm.
The mechanism of up-conversion fluorescence emission of CDs is due to
a two-photon absorption process. The recycling of CDs from coal
burning dust from coal-fired power plants are not only good to protect
environment but also would be helpful for mass production of CDs.
\end{abstract}
\maketitle

\parindent=0cm
\section*{1. Introduction}

Carbon dots (CDs) are zero-dimensional (0D) carbon-based nanomaterials
discovered in 2004 \cite{Xu04}. They are composed of sp$^2$/sp$^3$ carbon
skeletons and a variety of surface functional groups such as hydroxyl,
carboxyl, carbonyl, and amino groups. The sizes of CDs are generally
less than 10 nm and their cores usually consist of aromatic structures
with sp$^2$ hybridized carbon \cite{Li22}. CDs exhibit high photoluminescent
quantum yield and luminescence stability, along with good biocompatibility,
low toxicity, excellent chemical inertia, low production cost, abundant raw
materials, and environmental friendliness. These characteristics make CDs
one of the emerging fluorescent materials poised to replace traditional organic
fluorescent dyes and semiconductor quantum dots \cite{Yu21,Wang17}.
Nowadays, CDs have attracted significant attention and shall be widely used
in many fields such as optoelectronic devices, green lighting, biological
imaging, cell labeling, drug delivery, cancer diagnosis and treatment,
etc \cite{Li22,Akbar21,Hu16,Li17,Zhang19,Baker10,Zhou13}.

Coal and coal-based derivatives are known for their high carbon content,
which is rich in aromatic structures and graphite microcrystalline. The organic
part of coal contains aromatic and aliphatic compounds, with aromatic
compounds forming the basic skeleton of coal \cite{Zhu19}. Additionally, coal contains a
large number of polycyclic aromatic hydrocarbons (PAHs), primarily derived
from coal-forming precursors such as plant remains and microorganisms. Many
research groups have explored using coal as a carbon source to prepare high-quality
CDs due to coals aromatic structure and abundant
resources \cite{Hu16,Chu22,Li17,Sasikala16,Zhang19ACS,Kovalchuk15,
Thiyagarajan16,Feng19,Geng17,Singamaneni15,Dong12,Hu14,Hu17,Zhang17,
Kang19}. The unique aromatic structure in coal offers advantages such as
abundant raw materials and low cost compared to pure sp$^2$ carbon allotropes
like graphite for the preparation of CDs \cite{Ye13}. In 2016, Hu et al.
proposed a facile, green, and inexpensive top-down strategy for producing
fluorescent CDs from coal, avoiding the burden of tedious or inefficient
post-processing steps and the risk of highly toxic gas release \cite{Hu16}.

At present, coal is one of the most abundant energy resources in nature
and is typically used as the main fuel for thermal power generation worldwide.
During electricity generation from coal combustion, a mixture of flue gases
is released, including CO, CO$_2$, SO$_2$, SO$_3$, NO, NO$_2$, toxic organic
pollutant like PAHs, and particulate matter (PM) of various sizes, such as
carbon black \cite{Anderson02,Choi12}. The dust emitted from coal-fired power
plants contains water-soluble carbon dots (CDs), water-insoluble black carbon,
and inorganic components such as sulfates. However, current flue gas purification
technologies are difficult to capture and control CDs with the size of a few nanometers.
As a result, these particles are discharged into the atmosphere, causing
environmental pollution without effective monitoring \cite{Din13,Highwood16}.
The emission of particulate matter from coal burning is substantial due to the
high number of coal-fired power plants. Because of their small particle size,
large specific surface area, and high reactivity, CDs can easily adsorb heavy
metals, microorganisms, and other toxic substances, which can directly enter
the human alveoli and harm human health.

In previous studies, the precursors of CDs have mainly focused on various
types of biomass and chemical reagents. The separation and extraction of CDs
from coal-burning dust in coal fired power plants are rarely reported yet.
In this study, CDs were separated and extracted from the smoke washing waste
water obtained from coal-burning dust in a coal fired power plant. The formation
mechanism, structural properties, and up-conversion and down-conversion
fluorescence emission mechanisms of the CDs were examined. These CDs are of a
few nanometers in size and exhibit good water solubility. Typically, coal burning
dust are discharged into the atmosphere, which would contribute significantly
to PM$_{2.5}$ (sizes of PM $<$2.5 $\mu m$) pollution because they contains large amounts
of CDs with size of a few nanometers. Therefore, the investigating the formation
mechanism and fluorescence properties of CDs in coal burning dust from coal-fired
power plants holds significant research value and practical implications for resource
utilization, environmental protection, and carbon neutrality. Meanwhile, there can also
find a low-cost, non-polluting, and sustainable method to produce CDs using coal
dust as a raw material.

\section*{2. Results and discussions}
\label{Results and discussions}

\textbf{Figure 1} illustrates the primary mechanisms of CDs formation during coal
combustion. As shown in \textbf{Figure 1}(a), PAHs condense and polymerize to
form CDs with a carbon skeleton sheet created through sp$^2$ hybridization.
The self-assembly of PAHs, either originally present in coal or generated from
its incomplete combustion, is one of the main mechanisms of CD formation during
coal combustion \cite{Xue23,Shi23}.
Another key formation mechanism of CDs occurs when coal burns at high temperatures
of 700 $-$ 1200 $^\circ$C with breaking the bridge bonds which link the edges of
microcrystalline fragments in coals \cite{Hu16,Zhu19}.
This process results in the formation of graphene fragments with functional
groups at their edges, known as graphene quantum dots (GQDs) or CDs as depicted
in \textbf{Figure 1}(b).

\begin{figure}
  \includegraphics[width=10cm]{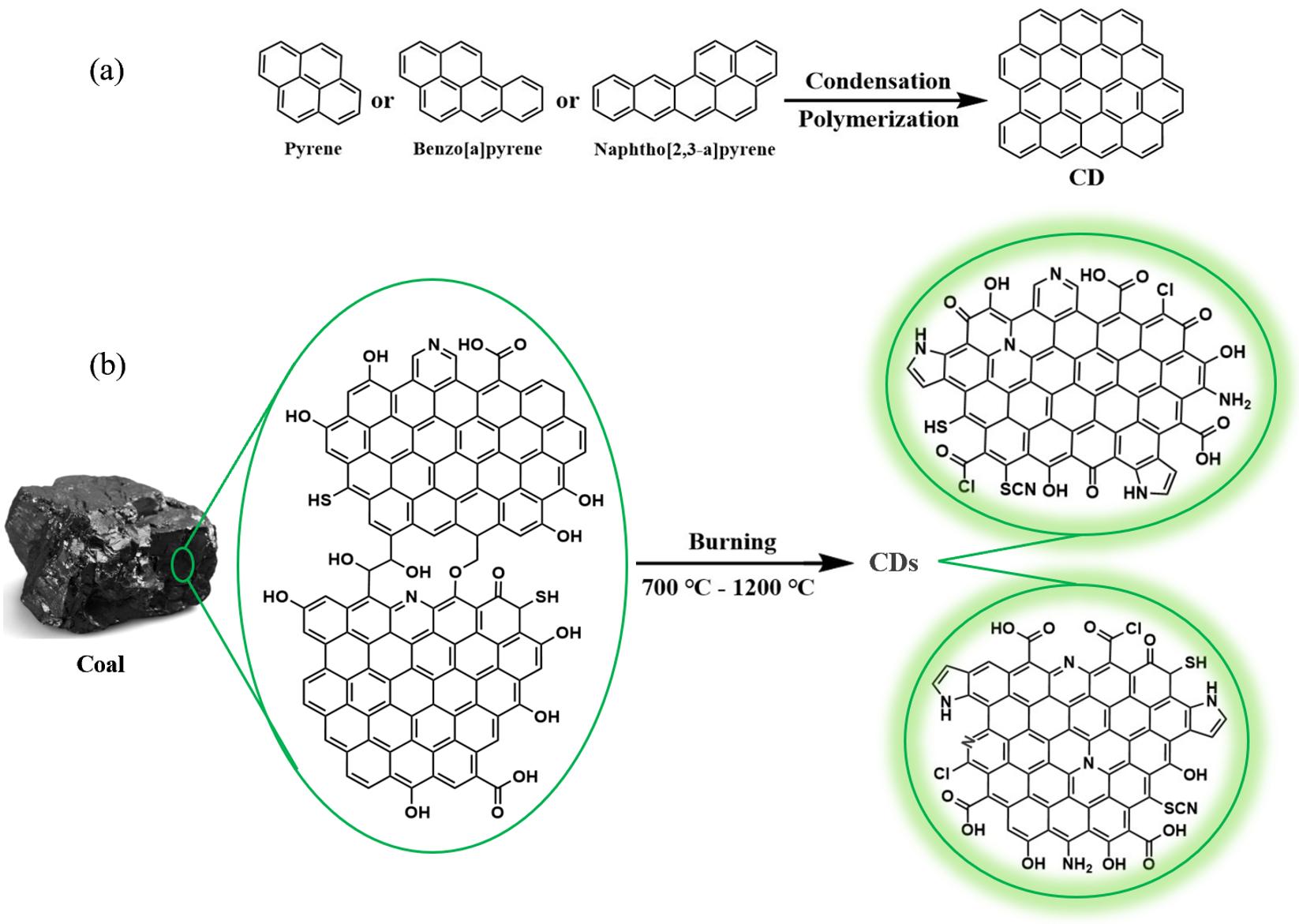}
  \caption{Schematic diagram of the possible main mechanisms of CDs
  formation during coal combustion. (a) The CD formation
  is due to condensation and polymerization of PAHs during coal combustion.
  (b) The bridge bonds linking edges of aromatic structures or microcrystalline
  fragments in coal are breaking and forming CDs with different functional
  groups when the coals are burning.}
\end{figure}

\begin{figure}
  \includegraphics[width=10cm]{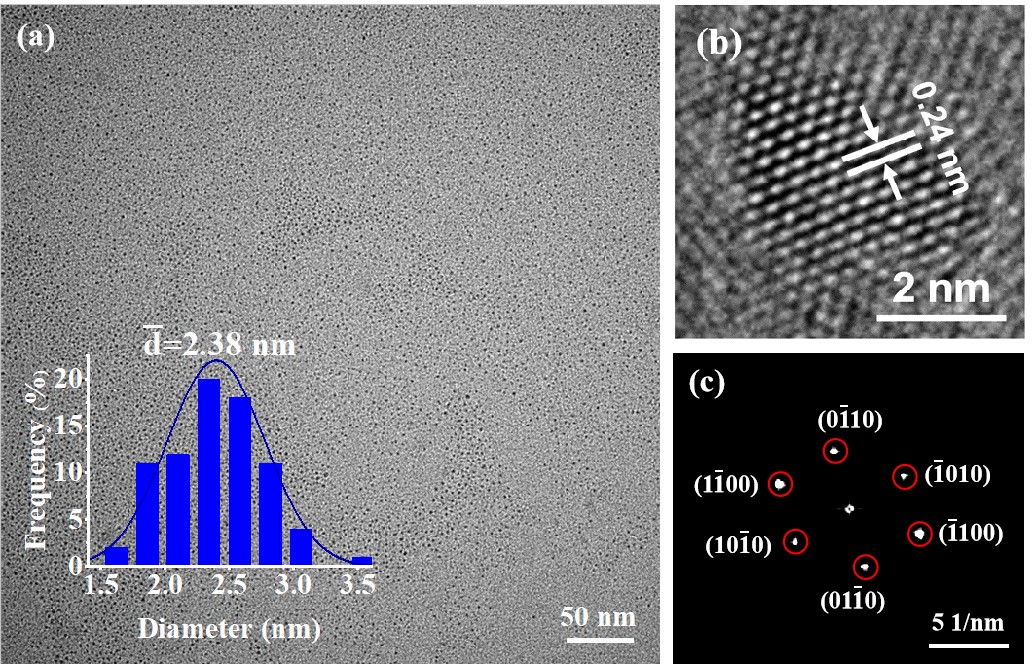}
  \caption{(a) TEM image, (b) HRTEM image, and (c) fast fourier transform
  (FFT) diffraction pattern of CDs extracted from coal power plant smoke
  washing wastewater in a coal fired power plant. The inset in (a) is the
  particle size distribution curve of CDs extracted from smoke washing wastewater.}
\end{figure}

\begin{figure}
  \includegraphics[width=8.6cm]{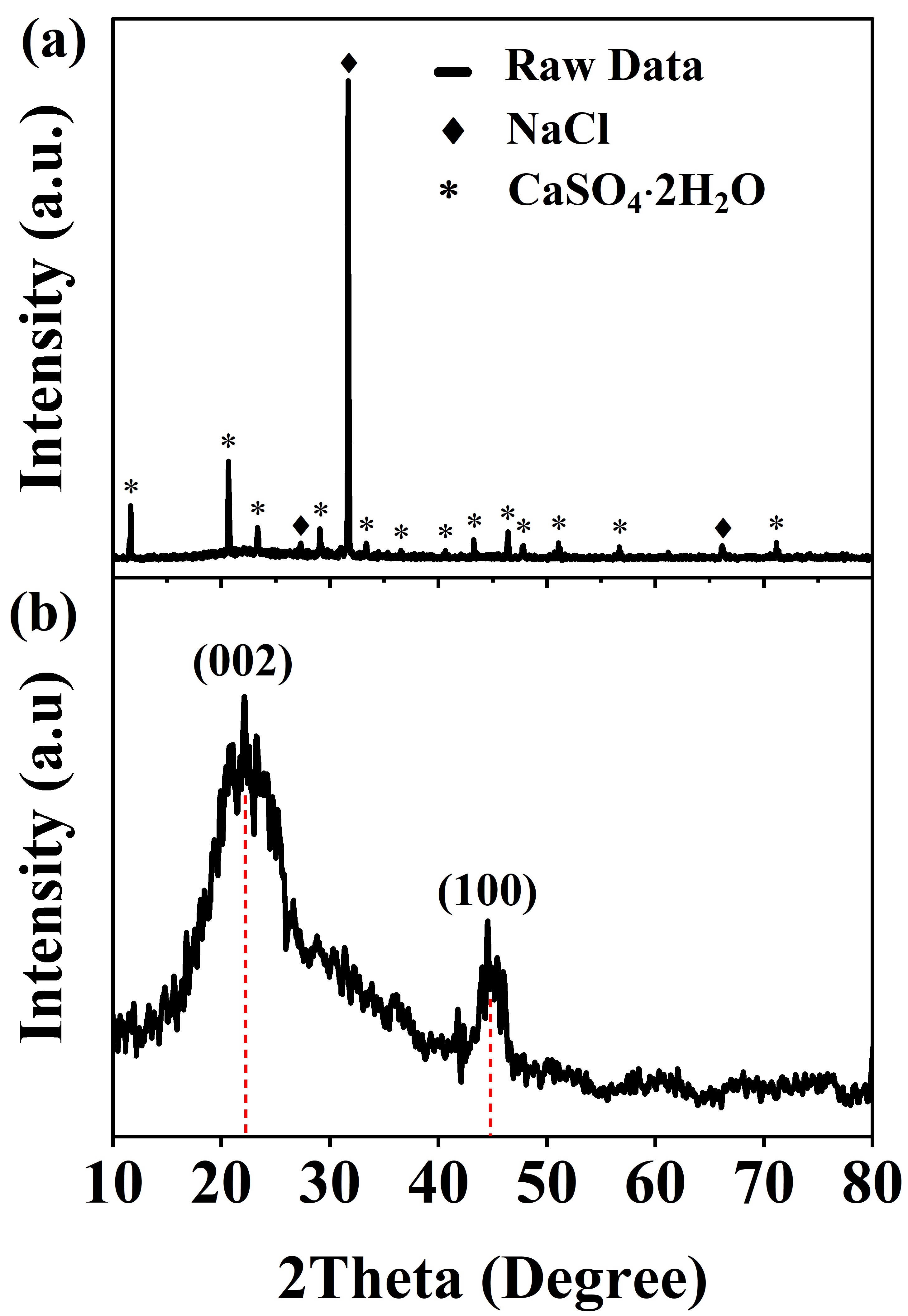}
  \caption{(a) XRD pattern of inorganic impurity crystals precipitated from
  CDs solution separated and extracted from smoke washing wastewater of
  a coal fired power plant. (b) XRD pattern of CDs after purification.}
\end{figure}

\begin{figure}
  \includegraphics[width=8.6cm]{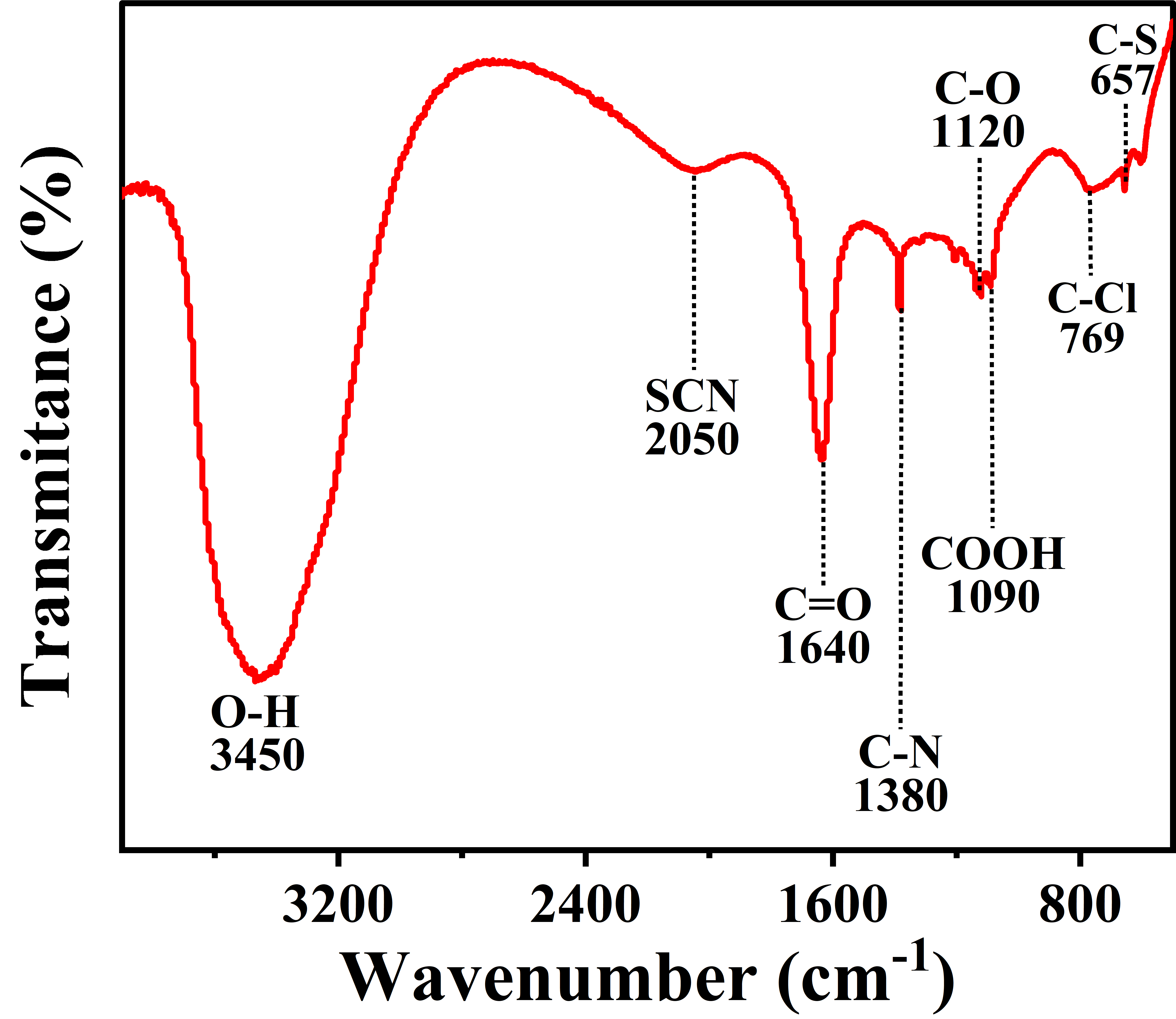}
  \caption{FT-IR spectra of CDs extracted from smoke washing wastewater of
  a coal fired power plant.}
\end{figure}

\textbf{Figure 2} presents the TEM images which show the morphology and
lattice fringes, as well as the corresponding diffraction pattern of CDs
extracted from smoke washing wastewater discharged from the wet flue gas
desulfurization system in the absorption tower of a coal fired power plant.
The inset of \textbf{Figure 2}(a) shows that the size of CDs ranges from
1.5 nm to 3.5 nm, with an average diameter of 2.38 nm. \textbf{Figure 2}(b)
clearly displays a crystal plane spacing of 0.24 nm in the conjugated sp$^2$-domain
of CDs, corresponding to the (1120) crystal plane of graphene \cite{Qu14}. The
hexagonal diffraction pattern in \textbf{Figure 2}(c) corresponds to the
six symmetrical crystal faces of the graphene sheets or graphite microcrystalline
at the core of CDs, indicating aromatic structures in CDs with high crystallinity.

\textbf{Figure 3} shows X-ray diffraction (XRD) patterns of (a) CDs
with impurities and (b) purified CDs solution. \textbf{Figure 3}(a)
reveals that the diffraction peaks of impurities in the CD solution
correspond to NaCl (JCPDS file no. 99-0059) and
CaSO$_4$$\cdot$2H$_2$O (JCPDS file no.74-1904) crystals.
\textbf{Figure 3}(b) displays two typical peaks at 22.04$^\circ$
and 44.58$^\circ$, corresponding to the (002) and (100) crystal planes
of the graphitic structure, respectively \cite{Li22}.
The mechanism for the separation of CaSO$_4$$\cdot$2H$_2$O crystals in
the CD solution is as follows
\begin{equation}
\begin{split}
        Absorption:&SO_2+H_2O \rightarrow H_2SO_3,\\
        Neutralization:&CaCO_3+H_2SO_3 \rightarrow CaSO_3+CO_2\\
        Oxidation:&CaCO_3+1/2O_2 \rightarrow CaSO_4\\
        Crystallization:&CaCO_3+1/2H_2O \rightarrow CaSO_3\cdot1/2H_2O\\
        Crystallization:&CaCO_4+2H_2O \rightarrow CaSO_4\cdot2H_2O
\end{split}
\end{equation}

The NaCl and CaSO$_4$$\cdot$2H$_2$O crystals in CDs solution were
removed by washing with distilled water several times \cite{Teng14,Chen17}.
The masses of NaCl and CaSO$_4$$\cdot$2H$_2$O crystals separated
from CDs solution were measured using a precision balance.
The combined mass of these crystals accounted for approximately
56.5\% of the total mass of the CDs solution.
Trace elements in the CD solution were measured using inductively
coupled plasma (ICP)spectroscopy. The result shows that the CD
solution contained 373.50 mg/L calcium, 394.23 mg/L sulfur, 44.32 mg/L sodium,
2.96 $\mu$g/L iron, and 5.96 $\mu$g/L of Plumbum. The mass ratio
of calcium to sulfur is close to 1:1 with calcium being slightly
less than sulfur. However, the molar ratio of Ca to S in CaSO$_4$$\cdot$2H$_2$O
crystal should be 1:1, implying a mass ratio of 5:4. This suggests
the presence of a small amount of sulfite (H$_2$SO$_3$) in the CD
solution, as indicated by the precipitation process in Equation 1.
The pH of the CD solution is 6.4, with acidity arising from trace
sulfite and nitric acid produced by the chemical reactions of
O$_2$ and NO$_x$ in the coal flue gas.

The Fourier-transform infrared (FT-IR) spectrum of the CDs is shown in \textbf{Figure 4}.
The broad peak around 3456 cm$^{-1}$ corresponds to the stretching vibration of O-H \cite{Li22}.
The absorption at 2050 cm$^{-1}$ is due to the stretching vibration of the -SCN group \cite{Miao17}.
The strong peak at 1640 cm$^{-1}$ is attributed to the stretching vibration of C=O \cite{Zhu17}.
The peak around 1380 cm$^{-1}$ corresponds to the C-N stretching vibration \cite{Li22}.
The C-O stretching vibration appears at about 1120 cm$^{-1}$ \cite{Liu19}, and the absorption near 1090 cm$^{-1}$ is due to the -COOH group \cite{Dang19}. Absorptions at 760 cm$^{-1}$ and 655 cm$^{-1}$ are attributed to the stretching vibrations of C-Cl and C-S, respectively \cite{Zhang21J,Hu20N}.

\begin{figure}
  \includegraphics[width=16cm]{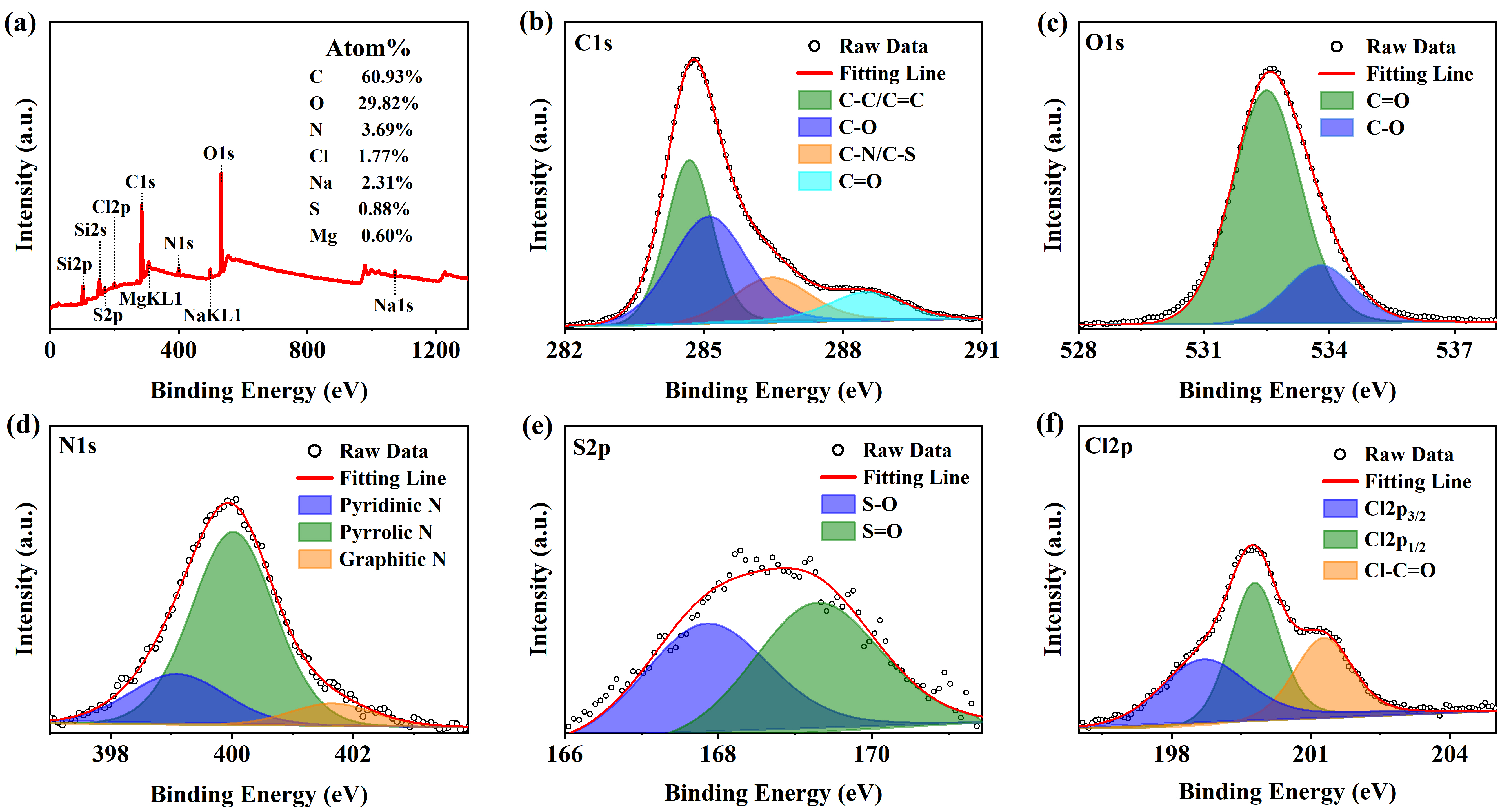}
  \caption{XPS spectra of CDs separated and extracted from smoke washing
  wastewater of a coal fired power plant.}
\end{figure}

\begin{figure}
  \includegraphics[width=8.6cm]{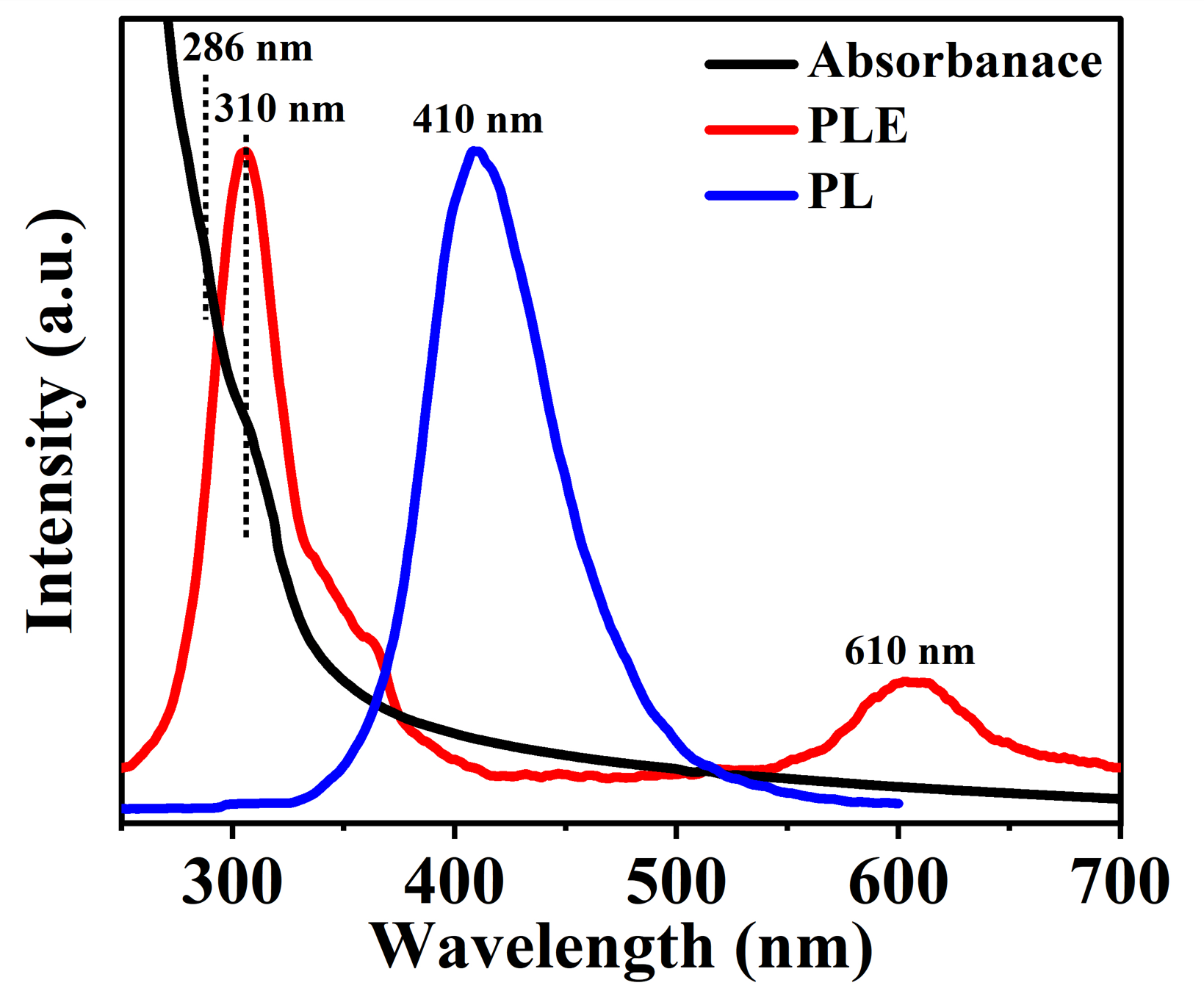}
  \caption{Absorption spectrum (ABS), photoluminescence excitation
  spectrum (PLE), and photoluminescence spectrum (PL) of
  CDs separated and extracted from smoke washing wastewater of a
  coal fired power plant.}
\end{figure}

\begin{figure}
  \includegraphics[width=10cm]{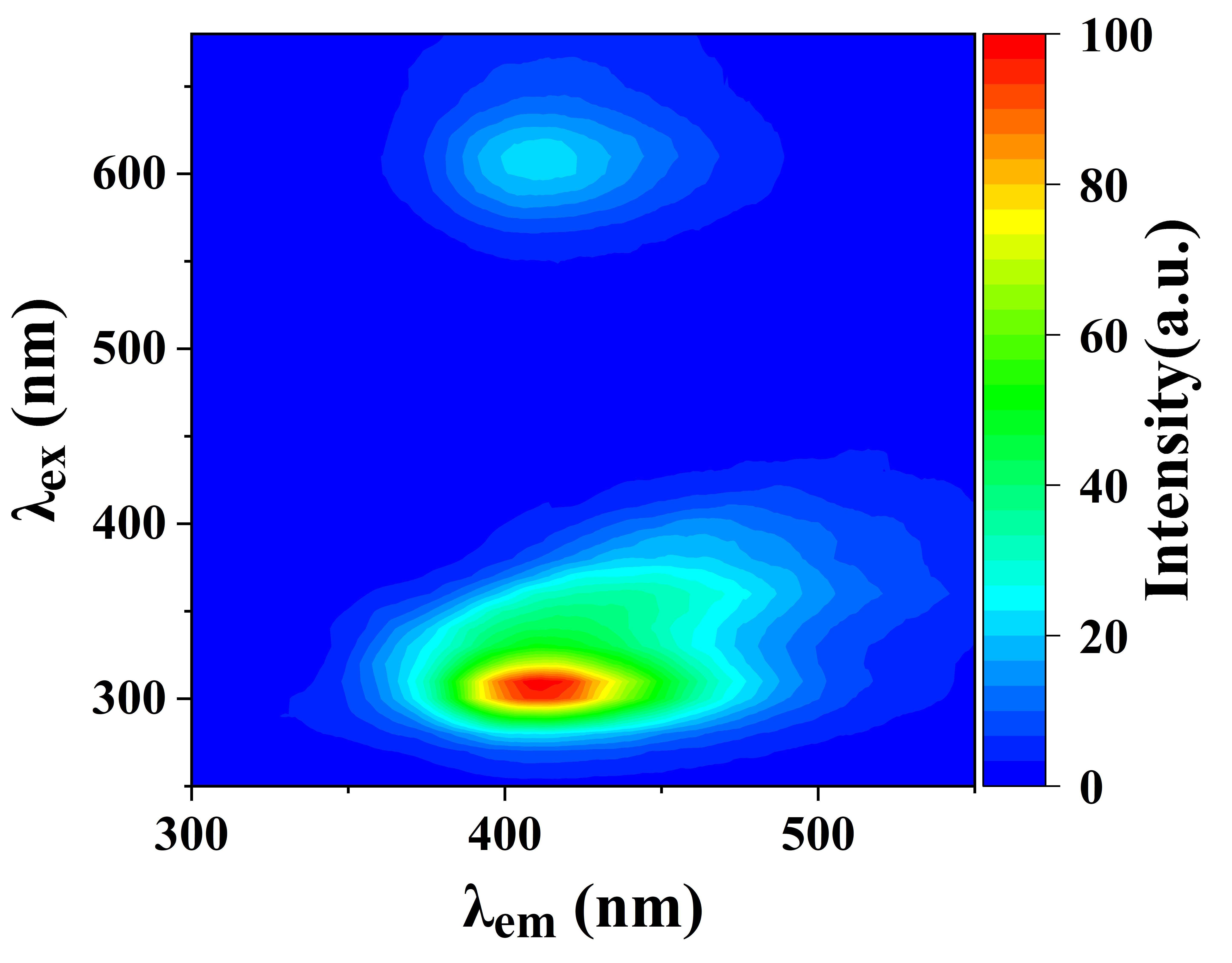}
  \caption{Excitation and emission contour map of CDs separated and extracted
  from smoke washing wastewater of a coal fired power plant.}
\end{figure}

\begin{figure}
  \includegraphics[width=16.4cm]{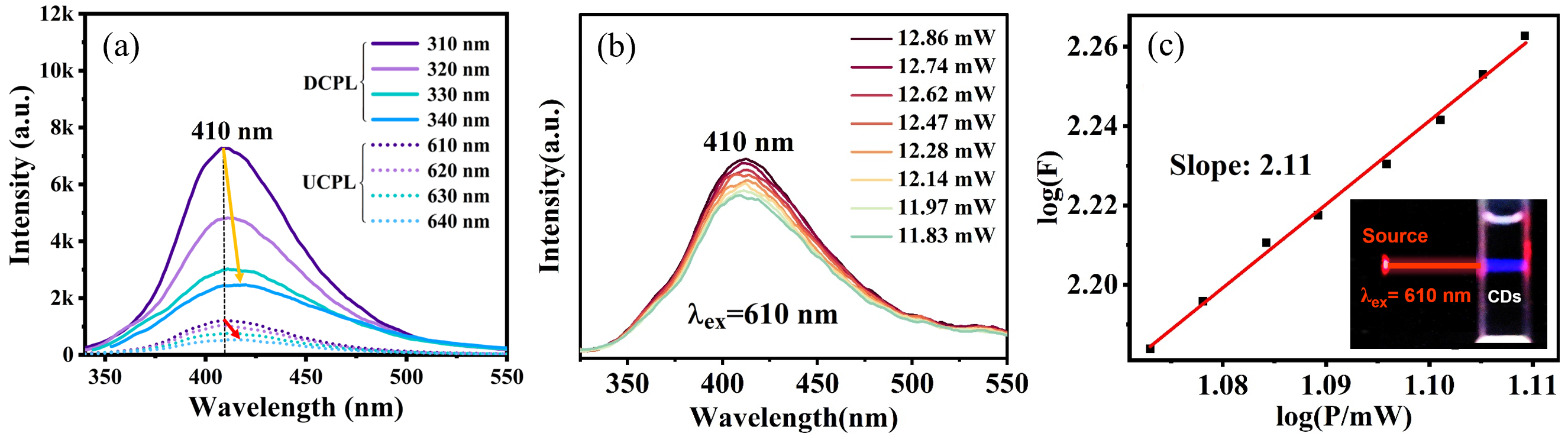}
  \caption{(a) DCPL and UCPL spectra of CDs separated and extracted from
  smoke washing wastewater of a coal fired power plant. (b) PL spectra of
  CDs with different excitation light powers. (c) Logarithmic plots of the
  dependence of two photon excitation induced fluorescence of CDs on the
  irradiation power of excitation light at 610 nm. The fitted slope is 2.11
  and the estimated uncertainty of the fitted slope is $\pm$0.03. The inset
  in (c) is the photograph of CDs emitting purple-blue up-conversion fluorescence
  when excited by the red light with a wavelength of 610 nm.}
\end{figure}

The X-ray photoelectron spectroscopy (XPS) analysis of the CDs is shown in \textbf{Figure 5}.
The CDs are primarily composed of C(60.93\%), O(29.82\%), and N(3.69\%), with small amounts
of impurities such as Na, Cl, S, and Mg. The C1s peaks can be divided into four peaks
corresponding to the bonds
C-C/C=C (284.7eV, 36.47\%), C-O (285.1eV, 38.35\%), C-O/C-S(286.5eV, 15.49\%),
and C=O (288.5eV, 9.69\%), respectively \cite{Wei19}. The O1s peaks can be divided
into two peaks corresponding to the bonds C=O (532.2eV, 80.38\%) and
C-O (533.8eV, 19.62\%) \cite{Yang22,Liu22}. The N1s peaks can be divided into
three peaks corresponding to pyridinic N (399.1eV, 21.43\%),
pyrrolic N (400.0eV, 70.37\%), and graphitic N-C$_3$ (401.7eV, 8.19\%) peaks \cite{Sun13,Ding14}.
The Cl2p can be divided into three peaks: Cl $^2$p$_{3/2}$ (198.5 eV, 26.09\%),
Cl $^2$p$_{1/2}$ (199.8 eV, 48.52\%) and Cl-C =O (201.3 eV, 25.39\%) \cite{Zhu20}.
The S2p peak can be divided into two peaks corresponding to S-O (167.8 eV, 47.97\%) and
S=O (169.3 eV, 52.03\%) \cite{Wang19NH}. The XPS results for impurities and
functional groups in the CDs are consistent with the FT-IR results. Impurity
elements such as Na, Cl, S, and Mg mainly originate from the coal. The Si
element comes from the silicon substrate used to place the sample during the
XPS measurement and is therefore excluded from the elemental composition of the CDs.

The black line in \textbf{Figure 6} shows the ultraviolet-visible (UV-vis)
absorption spectrum of the CDs. There are two typical wide absorption bands
near 286 nm and 310 nm, with the band near 286 nm attributed to the $\pi\rightarrow\pi^*$
transition of C=C, and and the absorption near 310 nm attributed to the $n\rightarrow\pi^*$
transition of C=O on the surface of CDs \cite{Li22}. With a fixed emission wavelength
of 410 nm and scanned the excitation wavelength, the photoluminescence excitation spectrum (PLE)
of CDs is given by the red curve in \textbf{Figure 6}. The PLE curve in \textbf{Figure 6}
indicates that CDs can emit fluorescence with a peak wavelength of about 410 nm when
excited by photons with wavelengthes both of 310 nm and 610 nm. The blue curves in
\textbf{Figure 6} is PL spectrum with a excitation wavelength of 310 nm. The PL and PLE
spectra reveal that the fluorescence around 410 nm has two excitation peaks which
a strong down-conversion excitation peak at 310 nm and a weak up-conversion excitation
peak at 610 nm.

In \textbf{Figure 7} we give the 2D contour plot of PL spectrum of CDs which
implies that fluorescence emission at wavelength 410 nm has two electronic transition channels.
One type of photoluminescence is down-conversion photoluminescence (DCPL), which corresponds
to $n\rightarrow\pi^*$ transitions of the C=O bond. Another type is up-conversion photoluminescence
(UCPL), which involves electronic transitions caused by continuous photon absorption coupled
with the stretching vibration of the C=O bond \cite{Dong15}. CDs contain a variety of impurities
or surface groups which makes them possessing a variety of surface states. Consequently,
such CDs are more likely to have up-conversion and down-conversion fluorescence
emissions due to the high probability of metastable states.

\textbf{Figure 8}(a) shows the up-conversion and down-conversion photoluminescence
(PL) spectra of CDs. As can be seen, CDs can emit purple-blue fluorescence
around 410-417 nm excited by red light with wavelengths of 610 nm, 620 nm, 630 nm,
and 640 nm, respectively. For the excitation wavelengths at 310 nm, 320 nm, 33 nm, and
340 nm, the photoluminescence peak is located around 410-420 nm. With the increasing
of excitation wavelength, the emission peak wavelengths of the up-conversion
photoluminescence (UCPL) and the down-conversion photoluminescence (DCPL)
have a red-shift of about 10 nm, as shown by the red arrow and the yellow arrow in
\textbf{Figure 8}(a). The red-shifts of UCPL and DCPL can be understand with the
help of \textbf{Figure 9}. Because CDs have multiple surface state levels,
the photon energy emitted from the electronic transitions from
different surface state levels to the ground state would be different.
It is interesting that the emission peak wavelengths for both UCPL and DCPL of
CDs are around 410 nm, which indicates that both UCPL and DCPL emit photons
through surface state luminescence at the same surface state energy levels.
\textbf{Figure 8}(b) reveals PL spectra of CDs with different excitation
light powers at a excitation wavelength of 610 nm. It's shown that the
fluorescence intensity of CDs increases slightly with increasing the excitation
light power. From \textbf{Figure 8}(c) we can get an obvious quadratic
linear relationship between the excitation power and the emission fluorescence
intensity of CDs while the slope is about 2.11, which can clearly indicating
that the UCPL mechanism of CDs is induced by two-photon excitation (TPE) \cite{Wang14,Jiang20}.
Through the inset in \textbf{Figure 8}(c), we can clearly see the UCPL
phenomenon in CDs.

The mechanisms of DCPL and UCPL can be understood with the help of \textbf{Figure 9}.
For the DCPL case, when the CDs are illuminating by photons with wavelengths of
310-340 nm (photon energy of about 4 eV), the electrons on C=O bonds would occur
$n\rightarrow\pi^*$ transitions as shown by the blue arrows.
Subsequently, the excited electrons move to the surface state energy levels
through non-radiative processes represented by dashed black arrows and combine
with holes in the ground state (n) with emitting the blue-purple fluorescence
with wavelengthes of 410-420 nm. For the case of UCPL, when CDs are irradiated
by the light with wavelengthes of 610-640 nm (with photon energies around 2 eV),
the electron in the ground state (n) absorbs a photon and jumps up to one of intermediate
energy levels induced by impurities or surface groups of CDs shown by black dashed lines, 
and then the electron in intermediate energy
levels could absorb another photon with the same
energy and transit to $\pi^*$ band \cite{Dong15}. Subsequently, the electron moves
to the surface state levels with a non-radiative transition process represented
by dashed black arrows and recombine with holes on the ground state (n) with emitting
fluorescence around 410-420 nm. This process can be regarded as two step photon
absorption from the ground state to the excited state through a virtual middle
energy level. Moreover, DCPL is more
stronger than UCPL because the probability of single-photon absorption is
greater than two photon absorption \cite{Jiang20,Wang14} in CDs .

\begin{figure}
  \includegraphics[width=10cm]{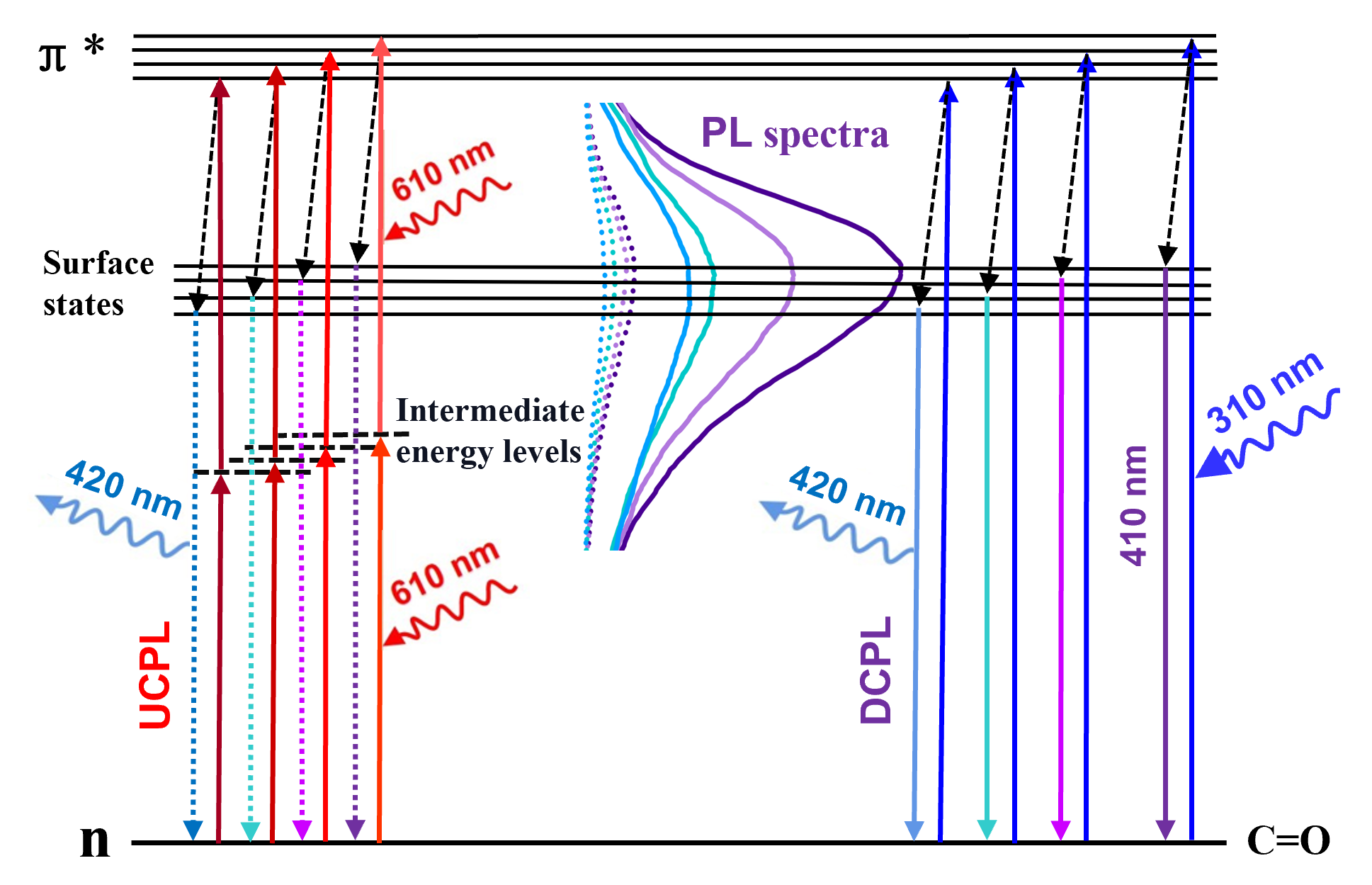}
  \caption{Schematic diagram of fluorescence emission mechanisms of DCPL and
  UCPL of CDs separated and extracted from smoke washing wastewater of a coal
  fired power plant.}
\end{figure}

\begin{figure}
  \includegraphics[width=8.6cm]{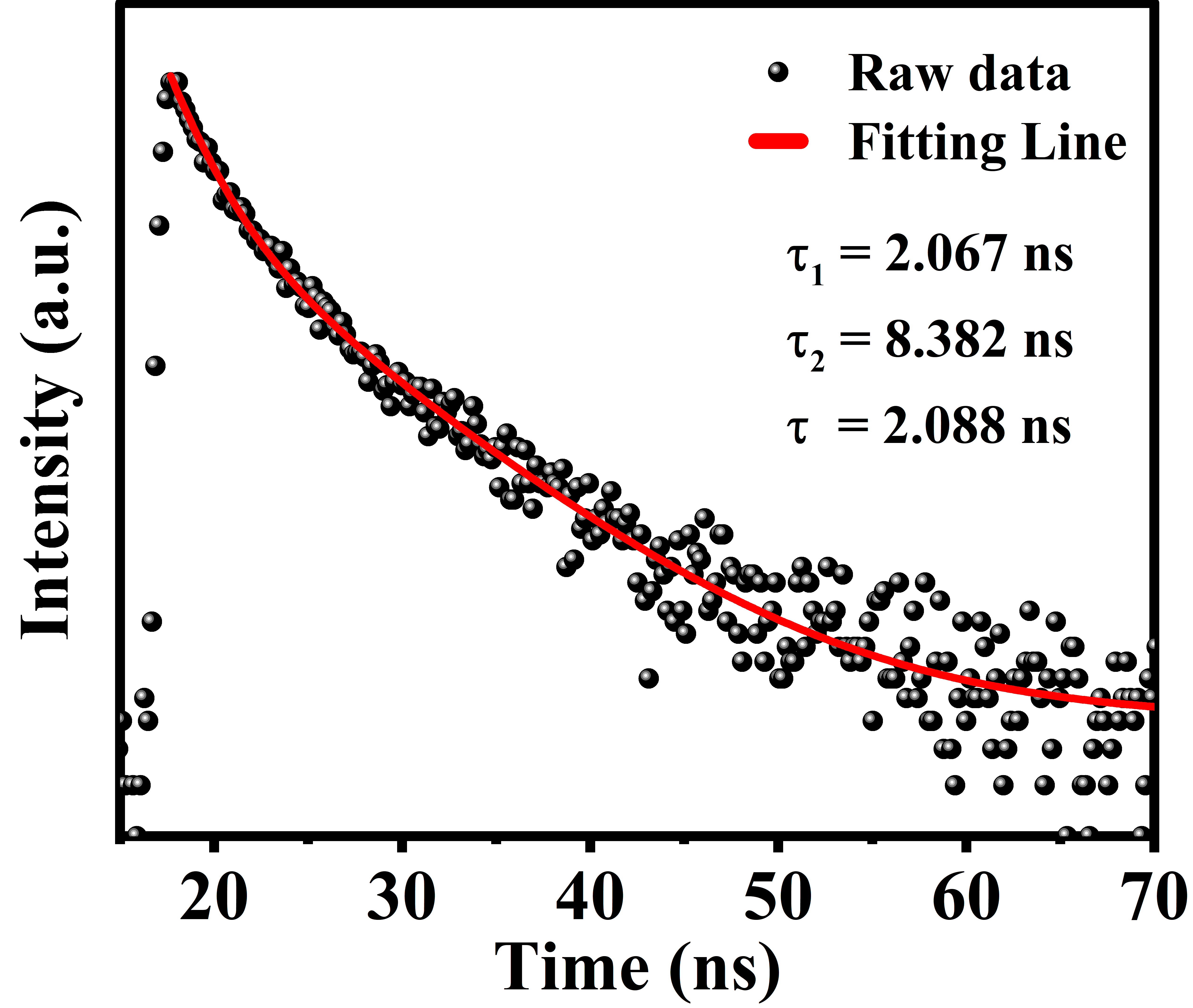}
  \caption{Lifetime decay curve of fluorescence CDs separated and extracted from smoke
  washing wastewater of a coal fired power plant.}
\end{figure}

\textbf{Calculation of fluorescence quantum yield (QY).} At a UV excitation wavelength of 310 nm, the fluorescence quantum yield of CDs can be calculated using quinine sulfate as the standard sample by using a combined FLSP fluorescence lifetime and steady-state fluorescence spectrometer equipped with integrating sphere (Edinburgh Instruments, UK). The formula for calculating the QY of fluorescence is given by
\begin{equation}\label{eq2}
Q=Q_{S}\cdot \frac{I_{S}}{I}\cdot \frac{A}{A_{S}}\cdot \frac{\eta^2}{\eta^2_{S}},
\end{equation}
where $Q_{S}$ is the quantum yield of the fluorescence for a standard sample for
reference. $I$ and $I_{S}$ are the integrated emission intensities of the CDs
sample and the standard sample, respectively. $A$ and $A_{S}$ are respectively the absorbance of the prepared
sample and standard sample at the same excitation wavelength. $\eta$ and $\eta_{S}$ are
respectively the refractivity of the prepared sample and standard sample.
It is calculated that the QY for DCPL of CDs is about 3.6\%.

The fluorescence decay life is further studied by measuring the
time-resolved fluorescence spectrum of CDs. The fluorescence
decay curve of CDs in \textbf{Figure 10} can be well fitted with two exponential
functions through
\begin{equation}\label{1}
R(t)=A_0+A_1e^{(-t/\tau_1)}+A_2e^{(-t/\tau_2)},
\end{equation}
where $A_0$ is the background PL intensity and $A_1$, and $A_2$,
are the fractional contributions to PL emission from
three transition channels with corresponding decay time or lifetime.

The average fluorescence lifetime
can be calculated through \cite{Dang16}
\begin{equation}\label{eq3}
\tau=\frac{A_1\tau_1^2+A_2\tau_2^2}{A_1\tau_1+A_2\tau_2}.
\end{equation}
According to the above Equation 4, the fluorescence lifetime of
CDs can be calculated as follows: $\tau_1$= 2.067 ns (99.9684\%),
$\tau_2$= 8.382 ns (0.0316\%), and the average fluorescence lifetime of CDs is
$\tau$= 2.088 ns.
The lifetime of 2.088 ns is attributed to the surface state in
CDs \cite{Yu16} as evidenced by the absorption of C=O in \textbf{Figure 4}.
The UCPL fluorescence mechanism of CDs is closely related to the functional
groups on their surfaces.

\section*{3. Conclusions}
\label{sec:conclusions}

In summary, we proposed a method to separate and extract CDs from
the coal-burning dust of coal-fired power plants and identified two
primary mechanisms for their formation. One mechanism involves the
self-assembly of PAHs contained in coal or produced by its incomplete
combustion, leading to the formation of CDs. Another mechanism involves
the breaking of bridge bonds linking different aromatic structures
in coal at high temperatures, resulting in CDs with various surface
functional groups. Our research demonstrates that these CDs have good
water solubility and exhibit up-conversion fluorescence behavior.
The up-conversion fluorescence mechanism is likely  a two-photon absorption
process. Both of up-conversion and down-conversion
processes can emit fluorescence at the same wavelength. CDs extracted
from coal-burning dust can emit purple-blue fluorescence at wavelengthes
around 410 nm when excited by red light with wavelengthes of 610-640 nm
or purple light with wavelengthes of 310-340 nm. The fluorescence quantum
yield at an excitation wavelength of 310 nm is about 3.6\%, and the average
fluorescence lifetime is $\tau =$ 2.088 ns which belongs to surface state luminescence.
This study indicates that CDs extracted from coal-burning dust possess
excellent fluorescence properties. Recycling CDs from coal-burning dust not
only helps protect the environment and turns waste into treasure but also
facilitates the mass production of CDs at a low cost and are
beneficial to achieve carbon neutrality.

\section*{4. Experimental Section}
Materials: CDs have good water solubility and most of them in the dust of coal fired
power generation are dissolved into the smoke washing wastewater discharged by the
dust cleaning system such as absorption tower. In this study, the smoke washing
wastewater discharged from wet flue gas desulphurization system in the absorption
tower from Yunnan Huadian Kunming Power Generation Co., Ltd. was used as raw material
to collect the CDs in the coal dust. The deionized water used in the experiment was
produced by Master-S15 deionized water production machine of Shanghai Hetai Instrument
Co., LTD.

Preparation: Coal-based CDs were separated and extracted from the smoke washing wastewater by the
procedures such as membrane filtration, centrifugation, dialysis, and concentration treatment.
The specific steps are as follows. Firstly, the smoke washing wastewater is
filtered through a multistage membrane to eliminate large impurity particles.
2000 mL filtered liquid is transferred to a beaker to heat at
70-80 $^\circ$C for 3-5 hours and then let it cool to room temperature.
The concentrated samples were dispersed in 30 mL deionized water.
Subsequently, the solution was put into an ultrasonic dispersion instrument
for oscillating 30 min and be centrifuged for 10 min by high speed centrifuge
at a speed of 12000 RPM. The supernatant was taken and filtered
by a 0.22 $\mu$m syringe filter membrane to obtain transparent
liquid. Then, this solution undergoes dialysis using MD44-500 Da membranes in water
for 24 hours.
Finally, the dialyzed solution was concentrated to 20 mL and the
CDs sample was obtained.

Characterization: The morphology and micro-structure of CDs were characterized by using
the transmission electron microscopy (Tecnai G2 TF30, USA) with 300 kV acceleration voltage.
The X-ray diffraction (XRD) patterns were obtained by Rotating target X-ray
polycrystalline diffractometer (RIGAKUTTRIII-18KW, Japan). The infrared absorption spectrum of CDs was measured by a Fourier
transform infrared spectrometer (Nicoletis10, USA).The X-ray photoelectron spectroscopy (XPS) of CDs were measured by
using PHI5000 Versa Probe II 51 photoelectron spectrometer with 1486.6 eV
Al K$_\alpha$. The UV-Vis absorption spectra were measured by a UV-Vis spectrophotometer (Specord200, Germany)
and the fluorescence spectrum of was measured by a fluorescence
spectrometer F9818012 (Shanghai lenguang technology Co., LTD). A combined fluorescence
lifetime and steady-state fluorescence spectrometer (Edinburgh
Instruments, UK) was used to measure the fluorescence decay curve at room temperature.

\section*{Acknowledgements}

This work was supported by the National Natural Science foundation
of China (NSFC) (Grants No. 12064049, No. 62175209, No. U2230122, No. U2067207,
No. 12364009, and No. 12004331), Yunnan Provincial
Science and Technology Department (Grants No. 202301AT070120, No. 202301AU070127),
Yunnan University (Grant No. CY22624108), Xingdian Talent Plans for Young Talents
of Yunnan Province (Grant No. XDYC-QNRC-2022-0492), and Shenzhen Science and Technology Program (Grant
No. KQTD20190929173954826).

\section*{Conflict of Interest}

The authors declare no conflict of interest.

\section*{Data Availability Statement}

The data that support the findings of this study are available from the
corresponding author upon reasonable request.

\section*{Keywords}
carbon dots, formation mechanism, fluorescence property, coal-fired power plants


\begin{thebibliography}{99}
\bibitem{Xu04}
X. Xu, R. Ray, Y. Gu, H. J. Ploehn, L. Gearheart, K. Raker, W. A. Scrivens, J. Am. Chem. Soc. \textbf{2004}, 126, 12736.

\bibitem{Li22}
P. Li, S. Xue, L. Sun, X. Zong, L. An, D. Qu, X. Wang, Z. Sun, Light Sci. Appl. \textbf{2022}, 11, 298.

\bibitem{Yu21}
J. Yu, X. Yong, Z. Tang, B. Yang, S. Lu, The Journal of Physical Chemistry Letters \textbf{2021}, 12, 7671.

\bibitem{Wang17}
Z. Wang, F. Yuan, X. Li, Y. Li, H. Zhong, L. Fan, S. Yang, Adv. Mater. \textbf{2017}, 29, 1702910.

\bibitem{Akbar21}
K. Akbar, E. Moretti, A. Vomiero, Adv. Opt. Mater. \textbf{2021}, 9, 2100532.

\bibitem{Hu16}
S. Hu, Z. Wei, Q. Chang, A. Trinchi, J. Yang, Appl. Surf. Sci. \textbf{2016}, 378, 402.

\bibitem{Li17}
M. Li, C. Yu, C. Hu, W. Yang, C. Zhao, S. Wang, M. Zhang, J. Zhao, X. Wang, J. Qiu, Chem. Eng. J. \textbf{2017}, 320, 570.

\bibitem{Zhang19}
Y. Zhang, K. Zhang, K. Jia, G. Liu, S. Ren, K. Li, X. Long, M. Li, J. Qiu, Fuel \textbf{2019}, 241, 646.

\bibitem{Baker10}
S. N. Baker, G. A. Baker, Angew. Chem. Int. Ed. \textbf{2010}, 49, 6726.

\bibitem{Zhou13}
L. Zhou, J. Geng, B. Liu, Part. Part. Syst. Charact. \textbf{2013}, 30, 1086.

\bibitem{Zhu19}
T. Zhu, R. Wang, N. Yi, W. Niu, L. Wang, Z. Xue, Int. J. Coal Sci. Technol. \textbf{2020}, 7, 19.

\bibitem{Chu22}
X. Chu, T. Chen, Y. Cao, Microchem. J. \textbf{2022}, 177, 107255.


\bibitem{Sasikala16}
S. P. Sasikala, L. Henry, G. Yesilbag Tonga, K. Huang, R. Das, B. Giroire,
S. Marre, V. M. Rotello, A. Penicaud, P. Poulin, ACS Nano \textbf{2016}, 10, 5293.

\bibitem{Zhang19ACS}
Y. Zhang, K. Li, S. Ren, Y. Dang, G. Liu, R. Zhang, K. Zhang, X. Long, K. Jia,
ACS Sustainable Chem. Eng. \textbf{2019}, 7, 9793.

\bibitem{Kovalchuk15}
A. Kovalchuk, K. Huang, C. Xiang, A. A. Mart\'{i}, J. M. Tour, ACS Appl. Mater. Interfaces \textbf{2015}, 7, 26063.

\bibitem{Thiyagarajan16}
S. K. Thiyagarajan, S. Raghupathy, D. Palanivel, K. Raji, P. Ramamurthy, Phys. Chem. Chem. Phys. \textbf{2016}, 18, 12065.


\bibitem{Feng19}
X. Feng, Y. Zhang, RSC Adv. \textbf{2019}, 9, 33789.

\bibitem{Geng17}
B. Geng, D. Yang, F. Zheng, C. Zhang, J. Zhan, Z. Li, D. Pan, L. Wang, New J. Chem. \textbf{2017}, 41, 14444.

\bibitem{Singamaneni15}
S. R. Singamaneni, J. van Tol, R. Ye, J. M. Tour, Appl. Phys. Lett. \textbf{2015}, 107.

\bibitem{Dong12}
Y. Dong, C. Chen, X. Zheng, L. Gao, Z. Cui, H. Yang, C. Guo, Y. Chi, C.M. Li, J. Mater. Chem. \textbf{2012}, 22, 8764.

\bibitem{Hu14}
C. Hu, C. Yu, M. Li, X. Wang, J. Yang, Z. Zhao, A. Eychm\"{u}ller, Y.P. Sun, J. Qiu, Small \textbf{2014}, 10, 4926.


\bibitem{Hu17}
S. Hu, X. Meng, F. Tian, W. Yang, N. Li, C. Xue, J. Yang, Q. Chang, J. Mater. Chem. C \textbf{2017}, 5, 9849.

\bibitem{Zhang17}
B. Zhang, H. Maimaiti, D.-D. Zhang, B. Xu, M. Wei, J. Photochem. Photobiol., A \textbf{2017}, 345, 54.


\bibitem{Kang19}
S. Kang, K.M. Kim, K. Jung, Y. Son, S. Mhin, J.H. Ryu, K.B. Shim, B. Lee, H. Han, T. Song, Sci. Rep. \textbf{2019}, 9, 4101.

\bibitem{Ye13}
R. Ye, C. Xiang, J. Lin, Z. Peng, K. Huang, Z. Yan, N.P. Cook, E. L. Samuel,
C.-C. Hwang, G. Ruan, Nat. Commun. \textbf{2013}, 4, 2943.


\bibitem{Anderson02}
R.R. Anderson, D.V. Martello, P.C. Rohar, B.R. Strazisar, J.P. Tamilia, K. Waldner, C.M. White, W.K. Modey, N.F. Mangelson, D.J. Eatough, Energy fuels \textbf{2002}, 16, 261.


\bibitem{Choi12}
J.-K. Choi, J.-B. Heo, S.-J. Ban, S.-M. Yi, K.-D. Zoh, Atmos. Environ. \textbf{2012}, 60, 583.

\bibitem{Din13}
S.A.M. Din, N.N.-H.N. Yahya, A. Abdullah, Pro-Soc Beh. Sci. \textbf{2013}, 85, 92.


\bibitem{Highwood16}
E.J. Highwood, R.P. Kinnersley, Environ. Int. \textbf{2006}, 32, 560.

\bibitem{Xue23}
S. Xue, P. Li, L. Sun, L. An, D. Qu, X. Wang, Z. Sun, Small \textbf{2023}, 19, 2206180.

\bibitem{Shi23}
L. Shi, B. Wang, S. Lu, Matter \textbf{2023}, 6, 728.

\bibitem{Qu14}
D. Qu, M. Zheng, L. Zhang, H. Zhao, Z. Xie, X. Jing, R.E. Haddad, H. Fan, Z. Sun, Sci. Rep. \textbf{2014}, 4, 5294.

\bibitem{Teng14}
X. Teng, C. Ma, C. Ge, M. Yan, J. Yang, Y. Zhang, P. C. Morais, H. Bi,
J. Mater. Chem. B, \textbf{2014}, 2, 4631.

\bibitem{Chen17}
J. Chen, X. Zhang, Y. Zhang, W. Wang, S. Li, Y. Wang, M. Hu, L. Liu, H. Bi,
Langmuir \textbf{2017}, 33, 10259.




\bibitem{Miao17}
X. Miao, X. Yan, D. Qu, D. Li, F.F. Tao, Z. Sun, ACS Appl. Mater. Interfaces \textbf{2017}, 9, 18549.


\bibitem{Zhu17}
J. Zhu, X. Bai, Y. Zhai, X. Chen, Y. Zhu, G. Pan, H. Zhang, B. Dong, H. Song, J. Mater. Chem. C \textbf{2017}, 5, 11416.

\bibitem{Liu19}
G. Liu, B. Li, Y. Liu, Y. Feng, D. Jia, Y. Zhou, Appl. Surf. Sci. \textbf{2019}, 487, 1167.


\bibitem{Dang19}
Q. Dang, Y. Sun, X. Wang, W. Zhu, Y. Chen, F. Liao, H. Huang, M. Shao, Appl. Catal., B \textbf{2019}, 257, 117905.


\bibitem{Zhang21J}
D. Zhang, D. Chao, C. Yu, Q. Zhu, S. Zhou, L. Tian, L. Zhou, The Journal of Physical Chemistry Letters \textbf{2021}, 12, 8939.


\bibitem{Hu20N}
Y. Hu, Z. Yang, X. Lu, J. Guo, R. Cheng, L. Zhu, C.-F. Wang, S. Chen, Nanoscale \textbf{2020}, 12, 5494.

\bibitem{Wei19}
Z. Wei, B. Wang, Y. Liu, Z. Liu, H. Zhang, S. Zhang, J. Chang, S. Lu, New J. Chem. \textbf{2019}, 43, 718.

\bibitem{Yang22}
X. Yang, L. Ai, J. Yu, G.I. Waterhouse, L. Sui, J. Ding, B. Zhang, X. Yong, S. Lu, Sci. Bull. \textbf{2022}, 67, 1450.

\bibitem{Liu22}
 J. Liu, T. Kong, H. M. Xiong, Adv. Mater. \textbf{2022}, 34, 2200152.


\bibitem{Sun13}
D. Sun, R. Ban, P.-H. Zhang, G.-H. Wu, J.-R. Zhang, J.-J. Zhu, Carbon \textbf{2013}, 64, 424.


\bibitem{Ding14}
H. Ding, J.-S. Wei, H.-M. Xiong, Nanoscale \textbf{2014}, 6, 13817.

\bibitem{Zhu20}
Z. Zhu, P. Yang, X. Li, M. Luo, W. Zhang, M. Chen, X. Zhou, Spectrochim. Acta, Part A \textbf{2020}, 227, 117659.

\bibitem{Wang19NH}
Q. Wang, S. Zhang, B. Wang, X. Yang, B. Zou, B. Yang, S. Lu, Nanoscale Horiz. \textbf{2019}, 4, 1227.

\bibitem{Dong15}
H. Dong, L. Sun, C. Yan, Chem. Soc. Rev. \textbf{2015}, 44(6):1608.


\bibitem{Jiang20}
L. Jiang, H. Ding, M. Xu, X. Hu, S. Li, M. Zhang, Q. Zhang, Q. Wang, S. Lu, Y. Tian, H. Bi, Small \textbf{2020}, 16, 2000680.

\bibitem{Wang14}
J. Wang, Z. Zhang, S. Zha, Y. Zhu, P. Wu, B. Ehrenberg, J. Y. Chen, Biomaterials \textbf{2014}, 35, 9372.

\bibitem{Dang16}
H. Dang, L.-K. Huang, Y. Zhang, C.-F. Wang, S. Chen, Ind. Eng. Chem. Res. \textbf{2016}, 55, 5335.

\bibitem{Yu16}
Y. Choi, B. Kang, J. Lee, S. Kim, G.T. Kim, H. Kang, B.R. Lee, H. Kim, S.-H. Shim, G. Lee, O.-H. Kwon, B.-S. Kim, Chem. Mater. \textbf{2016}, 28, 6840.
\end{thebibliography}
\end{document}